\documentclass[6pt, preprint2]{aastex}

\begin{document}

\title{Evidence that Quasars and Related Active Galaxies are Good Radio Standard Candles and that they are Likely to be a Lot Closer than their Redshifts Imply}

\author{M.B. Bell\altaffilmark{1}}
\altaffiltext{1}{Herzberg Institute of Astrophysics,
National Research Council of Canada, 100 Sussex Drive, Ottawa,
ON, Canada K1A 0R6;
morley.bell@nrc.gc.ca}

\begin{abstract}

For many years some astronomers have continued to argue, using redshift periodicities and quasar-galaxy associations, that quasars may be closer than their redshifts imply. Here, for the first time using raw radio data, I re-examine this question and find new evidence that supports this argument. Using VLBA flux densities and angular motions in jets, I show that the central engine of quasars and BL Lac objects appears to be a good radio standard candle. Using this information, relative distances are calculated and absolute radio distances are then obtained by referencing to a source whose true distance has been obtained using Cepheid variables. The results reveal that in this model most of the strong radio sources found in early surveys are nearer than 100 Mpc.
 For such nearby sources the measured redshifts must be almost entirely intrinsic, density evolution is unlikely and, if they are standard candles, their radio luminosity cannot vary significantly with intrinsic redshift, which should result in flat flux density vs redshift plots. LogN-logS plots must have a slope of -3/2, and this is expected to be true even over narrow intrinsic redshift ranges. Source number counts must also increase with distance as the cube. All of these requirements are shown to be met in the local model, confirming that it is a viable one. On the other hand, if the redshifts are assumed to be cosmological, none are met except the logN-logS slope of -3/2 which, because of evolution, is then \em not \em expected. When the data are fitted to the cosmological redshift (CR) model, the luminosity is found to increase while the spatial density decreases with redshift. 
However, the luminosity and density evolutions found turn out to be just \em those that would be created by the assumption that the redshifts are cosmological, when they are actually intrinsic. \em Thus the independent discovery here, using ejection velocities, that these sources are good radio standard candles, has led to the realization that the raw data fit the local model perfectly, and that the good logN-logS slopes likely occur because of this. The data can be fitted to the CR model, but not without making many additional, arbitrary assumptions. It is concluded that the flux density, and not the redshift, is likely to give the best indication of the distance to these objects. If so, they must then be a lot closer than previously thought.

\end{abstract}


\section{Introduction}

After \citet{sch63} and \citet{gre63} determined the redshift of 3C273 and 3C48 respectively, the redshifts of many more quasars were soon obtained. Most were quite large, and because these objects did not appear to be optical standard candles, generally becoming more luminous as their redshift increased \citep{bur67}, the question of the origin of their redshift immediately arose. However, since there appeared to be no way of explaining very large redshifts (z $>0.62$) \citep{bon64,buc59,buc66} other than by the expansion of the Universe, most astronomers soon concluded that the redshifts must be purely cosmological. Since that time most of the work done on quasars has attempted to show that the data supported that assumption and most quasar papers have assumed from the outset that the redshifts are cosmological.

However, a few astronomers have clung to the possibility that some, as yet unknown, mechanism might be responsible for at least a portion of the observed redshifts of these objects, and this belief has continued to be supported by evidence of possible periodicities in the redshift distributions \citep{kar71,kar77,bur01,bel04}, and by the apparent associations between high-redshift quasars and low-redshift galaxies \citep{arp97,arp98,arp99,bel02a,bel02b,bel02c,bel04,bur95,bur96,bur97a,bur97b,bur99,chu98,gal05,lop02}. Unfortunately, because of the almost immediate acceptance by most astronomers that quasar redshifts were cosmological, the normal course of scientific research has not been followed, and both sides of this issue have not been fairly examined.

In models where intrinsic redshifts are proposed, the current picture is that galaxies are born as QSOs that are ejected from the nuclei of active galaxies. At birth their redshifts contain a large intrinsic component of unknown nature, that gradually decreases. Their luminosity simultaneously increases and, after $\sim 10^{8}$ yrs, they mature into normal galaxies. Although the intrinsic component is very small in mature galaxies (similar to the peculiar velocity of the galaxy) it may never disappear completely, since small intrinsic components have also been found in the redshifts of mature spiral galaxies \citep{tif96,tif97,bel04a}. Since in this model the intrinsic redshift component continuously decreases, it has been referred to previously as the decreasing intrinsic redshift (DIR) model. A natural consequence of a decreasing, or age dependent, redshift component is that some radio galaxies will be more mature than others. Here I will refer to these as \em mature \em radio galaxies and \em young, active \em radio galaxies. It is assumed that the latter can have a significant intrinsic redshift component while this component in the former is almost negligible.

In the DIR model, as the objects age their optical luminosity increases, as their associated galactic component grows, and eventually they become mature galaxies that are bright enough to be detected to much larger distances. The small intrinsic redshift components found in galaxies have also been found to be superimposed on top of their distance, or cosmological, redshift component \citep[and references therein]{bel04a}. In this model it has not been necessary to give up the Big Bang, and this is one of the main ways that the DIR model differs from the decreasing intrinsic redshift model proposed by \citet{nar93}. 

It is suggested here that the evidence is now strong enough to warrant a re-examination of the question of intrinsic redshifts using a new approach.
Here, for the first time, I use accurate radio data to examine this question.
For this purpose I use the raw data found in radio surveys of quasars and active radio galaxies (angular motions, radio flux densities, source number counts, etc). The term \em raw data \em is defined here, as in Kellermann et al. (2004), as those observables that are uncontaminated by modeling. Because it is important not to influence the results by making unwarranted assumptions, the importance of using raw data cannot be emphasized strongly enough if a fair appraisal is to be obtained.

I first present evidence, in Section 2, indicating that flat spectrum quasars may be good radio standard candles and, as such, can be used as an independent check of quasar distances. It will be shown that, plotted on logarithmic scales, the upper envelope of the angular motions seen in flat-spectrum, $superluminal$ quasars and BL Lac objects increases with increasing 15 GHz flux density with the slope expected for a simple ejection model in which all ejection velocities are reasonably similar but are non-relativistic. A clearly defined upper envelope can be obtained only if the sources are good radio standard candles.
For standard candles, relative distances can be calculated using the flux density, and converted to absolute values by referencing to a source whose distance is known by some other means. It will be shown that this leads to distances less than 100 Mpc for these sources, and that they therefore occupy Euclidean space in this model.



Second, in Section 3, I examine two independent samples of extragalactic jetted radio sources. In the CR model the observations can only be explained if the radio luminosities of both the core and giant radio lobe components of these objects increase significantly with increasing redshift. The results also require that the luminosity increase be remarkably similar in both the core and lobe components.


Finally, in Section 4, having presented evidence in Section 2 that these sources are standard candles in Euclidean space, I examine logN-logS plots for four samples of quasars and active radio galaxies, and obtain slopes of -3/2 in all cases, confirming that, if the above is true, the sources must also be uniformly distributed. In the local model, where redshifts must be almost entirely intrinsic, when the source S vs z distribution is subdivided into several smaller intrinsic redshift ranges, good logN-logS slopes near -3/2 are still expected to be obtained for the sources in each narrow redshift range.


It is also demonstrated that the luminosity and density evolutions found in the CR model are exactly those predicted to be introduced by assuming incorrectly that the intrinsic redshift is cosmological.

Hereafter radio standard candles will be referred to simply as standard candles and optical standard candles will be referred to as optical standard candles. Note also that the term \em local model \em refers here to the fact that the sources being discussed are local in that model. They are nearby only because they were found in early radio surveys where the sensitivity was poor. It does not mean to imply that all sources in the local, or DIR, model are local. Finally, although the CR model is referred to here in general terms, because it has been discussed in detail by many others over the last 40 years it will not be discussed in detail here.


\section{Radio Distance}

In the CR model, anomalies were soon found related to rapid variability in extragalactic radio sources \citep{sho65,den65,pau66}, and later, to apparent superluminal ejection velocities \citep{coh77}. In simple ejection models, if source distances are known, changes in the angular positions of 'blobs' associated with their jets can be converted into linear ejection velocity components perpendicular to the line-of-sight. However, it has been known since the 1970s that this simple interpretation leads, in many cases, to apparent velocities $v_{app} = \beta_{app}c$, where $\beta_{app}$ is greater than 1 and c is the speed of light. The velocity thus appears to exceed the speed of light and has been referred to as \em superluminal. \em 
Although this apparent anomaly can be readily explained if the objects are really much closer than their redshifts imply, models involving relativistic expansion were developed to explain the rapid variabilty \citep{wol66,ree66,ree67}, so astronomers did not have to give up their belief that redshifts are reliable distance indicators. Relativistic beaming models also nicely explained the apparent superluminal ejection velocities. However, although relativistic motions in the ejected blobs can explain the apparent anomaly, even after forty years there is little convincing observational evidence that this is the correct interpretation. Although evidence has been claimed this usually raises more questions than it answers. There may well be relativistic effects in the cores of active galaxies, and this is not being questioned here. What is being questioned is whether or not the ejection velocities of the discrete radio blobs are relativistic.
It is important to keep in mind that the apparent \em anomalies \em mentioned above also have a common factor: they both involve the assumption that the redshifts of these objects are a good indication of their distance.

\subsection{Angular motions in radio jets}

The results of a 10-year VLBA program to study motions in radio jets at 15 GHz has recently been reported \citep{kel04} and these excellent measurements can now be used to re-examine some of the characteristics of these active radio sources. As noted above, I differentiate here between mature radio galaxies and young, active radio galaxies. The latter, since they are younger and may more closely resemble quasars, are still expected in the DIR model to contain a significant intrinsic redshift.

\begin{figure}
\hspace{-1.0cm}
\vspace{-1.6cm}
\epsscale{0.9}
\plotone{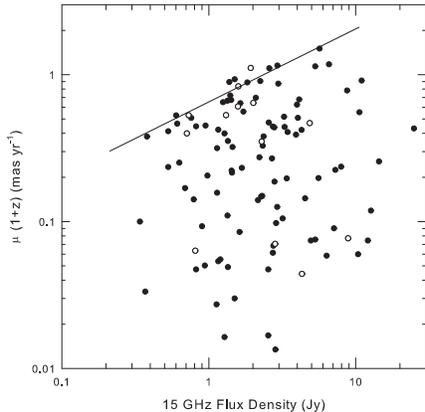}
\caption{{Angular motion $\mu$(1+z) (mas yr$^{-1}$) plotted versus radio flux density at 15 GHz for 109 quasars and BLLac objects from Table 2 of \citet{kel04}. The solid line has a slope of 1/2 and indicates that the flux density varies directly as the square of $\mu$(1+z), which is inversely related to distance in simple, non-relativistic ejection models. Thirteen quasars with less accurate motions plotted as open circles.  \label{fig1}}}
\end{figure}

\begin{figure}
\hspace{-1.0cm}
\vspace{-1.6cm}
\epsscale{0.9}
\plotone{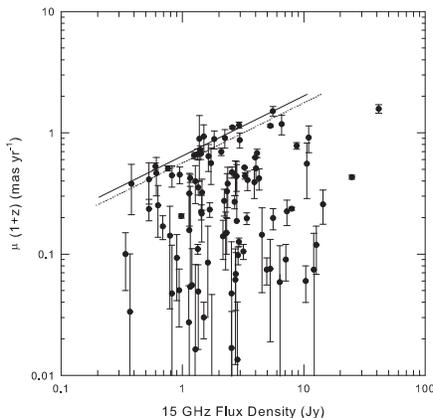}
\caption{{Same as Fig 1 with error bars. Thirteen quasars with less accurate motions have not been included. See text for a discussion of the open squares and dotted line. \label{fig2}}}
\end{figure}

Once a model is specified, there are several well defined relations that can be predicted.
If quasars are radio standard candles, their apparent radio power, or flux density, is also expected to fall off inversely as the square of their distance. When the flux density is plotted versus distance on logarithmic scales a slope of -2 will then be obtained.

In simple ejection models, if active radio galaxies and quasars are ejecting material at reasonably similar but non-relativistic velocities, the apparent angular motion of these 'blobs' per unit time will fall off linearly with distance, and it will also depend on the ejection angle relative to an observer. When plotted on logarithmic scales, the maximum value, or upper envelope, of these angular motions will fall off inversely with distance with a slope of -1. The angular motion then becomes a direct indication of distance, or a standard yardstick. In this simple model, the maximum velocity is only seen for those ejections in the plane of the sky, and ejections closer to the l-o-s will fall below this upper envelope.

In Fig 1 the maximum angular motion per unit time in the source reference frame, $\mu$(1+z), observed for 109 $superluminal$ quasars and BL Lac objects from \citet{kel04} is plotted versus the 15 GHz VLBA flux density on logarithmic scales. Here $\mu$ is the angular motion in mas yr$^{-1}$ and z is the relevant measured redshift. Although only 79 of 92 quasars were included in the previous analysis \citep{kel04}, here 91 of the 92 quasars are included. Thirteen were previously left out because their motions were less accurately determined. So that it may be clearly seen that the 13 quasars left out of the earlier work do not affect the results found here, they have been plotted in Fig 1 as open circles. Because six of the BL Lac objects have no redshifts, only 18 BL Lac objects can be included.

The 15 GHz flux density (S$_{VLBI}$) is the total flux density taken from the VLBA images \citep{kel04}. Most is assumed here to be associated with the central engine. In Fig 1, the upper envelope of the angular motion for these sources increases with the 15 GHz flux density with a slope of 1/2. In a simple ejection model $\mu$(1+z) is a measure of the relative distance to the object and the flux density then falls off as the square of the distance. This is what is expected in a simple model, if the sources are "standard candles" and the ejection velocities are $non-relativistic$ and reasonably similar in all sources. The measured angular motion is then due to geometry alone, with no time contraction, and with the largest angular motion corresponding to ejections in the plane of the sky. It seems unlikely that this upper envelope, with the correct slope, could be obtained by chance and this is discussed in more detail in Appendix A. It would also seem to be difficult to explain this result in the cosmological model, where relativistic ejection velocities are required and a simple geometrical explanation is ruled out.

Although it has been assumed here that the factor (1+z) is needed to adjust the angular motions to the source reference frame, since the nature of the intrinsic redshift proposed here is unknown there is no way of knowing whether this assumption is correct. If removing the (1 + z) factor destroyed the upper envelope it might be considered as a test of whether or not the intrinsic redshift behaves in a manner similar to other known redshifts. Unfortunately, the same slope is obtained if this factor is not included, and it is therefore not a meaningful test. Since this is a plot of angular motion vs flux density, with the high- and low-redshift sources scattered at random throughout the plot, it is therefore no surprise that the factor (1+z) does not affect the result. This conclusion is also relevant to other small corrections that are a function of redshift, such as the K-correction. The K-correction is assumed here to be small since these sources were selected because they had flat spectra.  These data, without the 13 quasars with large uncertainties, are re-plotted in Fig 2 with error bars and they show the same upper-envelope slope. Note that the plotted errors are equal to the listed errors in $\mu$ multiplied by the factor (1+z).
 
In Figs 1 and 2, the tight fit of the upper envelope to the line is a measure of how good a standard candle the source is. In this model only the sources near the upper boundary, which represents sources ejected in the plane of the sky, will have significance. In order to obtain a value for the slope of the upper envelope the following approach was taken. Three sources were chosen that were felt to best define the upper envelope. These sources (0736+017, 1219+096, 2200+420) were selected because, a) they lie along the upper envelope, b) they cover a wide range of Flux Densities (0.6 to 5.67 Jy), and, c) for sources at the upper boundary their angular motions ($\mu$) have the smallest uncertainties. When the logarithms of the fluxes densities and the angular motions of these objects are plotted using linear co-ordinates a slope of 1/2 is expected. Three cases were examined: 1) A linear regression was carried out assuming uniform errors on $\mu$(1+z). For this case the slope \em m \em obtained was \em m \em = 0.48 with a std error of 0.033. 2) A linear regression was carried out that was weighted according to the uncertainties in $\mu$(1+z). Here the slope obtained was \em m \em = 0.473 with one std error of 0.03. 3) A linear regression was also carried out which included the errors in $\mu$, but did not include the factor (1+z). In this case the slope obtained was \em m \em = 0.478 with one std error of 0.018. Inclusion of the factor (1+z) here makes little difference because the redshift of these objects is small.

Assuming that the sources are standard candles, using the flux density their relative distances were calculated. These are easily converted to absolute distances by fitting the calculated distance of a reference source to its distance established by some other means. In this case Virgo A was used as the distance standard and a value D$_{Virgo}$ = 18 Mpc was assumed. In Fig 3 the maximum angular motions per unit time in the source frame, $\mu$(1+z), are plotted versus the radio distances on logarithmic scales. The upper envelope falls off with the slope of -1, and these strong, jetted sources found in early radio surveys can be seen to be closer than 100 Mpc, and therefore the space they occupy can safely be assumed to be Euclidean. 

Changing the distance assumed for Virgo shifts the entire distribution in Fig 3 to higher or lower distances but does not alter the overall shape of the distribution, which is set by the relative values of the 15 GHz flux density. In Fig 3, where D$_{Virgo}$ = 18 Mpc is assumed, the upper envelope of the $superluminal$ sources (solid line) is displaced to a slightly higher distance than is found for the upper envelope of several radio galaxies (dotted line) whose distances are also thought to be accurately known and are plotted here as open circles. This may imply that the correct value of D$_{Virgo}$ is slightly less than 18 Mpc. In fact, excellent agreement is obtained between the upper envelopes of these radio galaxies and the quasars and BL Lac objects when the Cepheid distance to the Virgo cluster is assumed (D$_{Virgo}$ = 14.6 Mpc) \citep{fre01}. Finally, when new $\beta_{app}$ values are calculated using the distances determined here, they give ejection velocities for most sources that are now not even mildly relativistic. However, it is also important to keep in mind that the absolute distances calculated here are tied to the 15 GHz flux density of one source, and much more work is required before it can be concluded that Virgo is the best reference source to use to obtain absolute distances. The distances calculated here have not been included in a table because they are still subject to changes. However, major departures from the calculated distances are not anticipated.

\begin{figure}
\hspace{-1.0cm}
\vspace{-1.5cm}
\epsscale{0.8}
\plotone{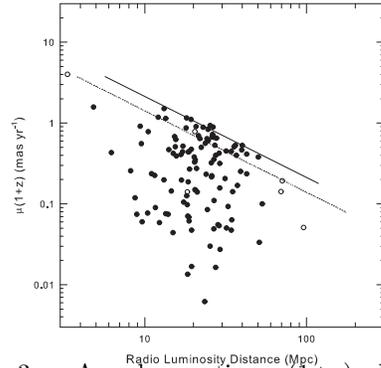}
\caption{{Angular motion $\mu$(1+z) plotted versus distance for quasars and BLLac objects, where distance has been determined by assuming that the sources are good  radio standard candle and the distance to Virgo is 18 Mpc. The upper envelope falls off with the expected slope of -1, as shown by the solid line. Open circles represent radio galaxies and have an upper envelope shown by the dotted line. A Hubble constant of H$_{o}$ = 58 is assumed. \label{fig3}}}
\end{figure}

\begin{figure}
\hspace{-1.0cm}
\vspace{-2.2cm}
\epsscale{0.8}
\plotone{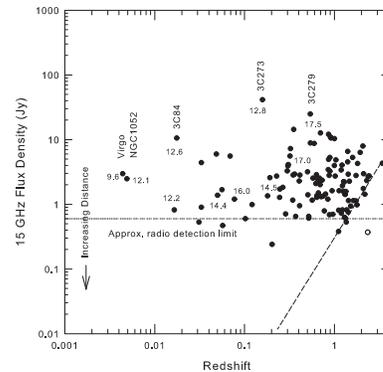}
\caption{{Radio flux density at 15 GHz plotted versus redshift for quasars, BLLac objects and radio galaxies with measured redshifts and angular motions from Table 2 of \citet{kel04}. The dotted line indicates the approximate radio detection limit, and the dashed line indicates a second clear cut-off at high redshifts. The numbers beside sources indicate their apparent magnitude. \label{fig4}}}
\end{figure}

\begin{figure}
\hspace{-1.0cm}
\vspace{-2.0cm}
\epsscale{1.0}
\plotone{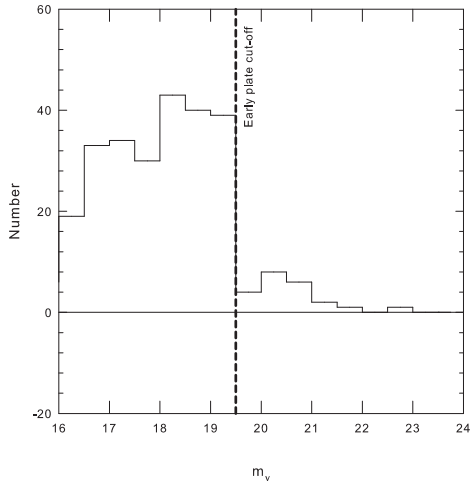}
\caption{{Distribution of optical apparent magnitudes for all quasars and BL Lac objects with measured redshifts in the \citet{sti94} sample.  \label{fig5}}}
\end{figure}

\begin{figure}
\hspace{-1.0cm}
\vspace{-1.7cm}
\epsscale{1.0}
\plotone{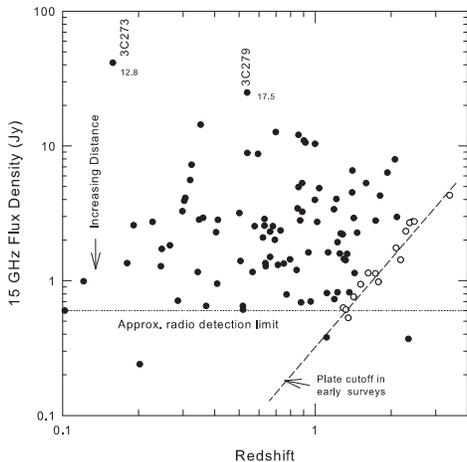}
\caption{{Same as Fig 4, only on an expanded scale. Sources defining the cut-off at the dashed line are indicated by open circles and their parameters are listed in Table 1. No redshift corrections (ie K-corrections, etc) have been applied to the data.   \label{fig6}}}
\end{figure}

\subsection{Selection Effects in the Data}

There are several selection effects likely to be present in the data. One is due to a radio detection limit, which is visible in the S$_{VLBI}$ 15 GHz data but is due to the 1 Jy radio limit set by the 5 GHz survey \citep{sti94} from which this sample of sources was selected. Another is the plate limit set by early optical work at apparent magnitude value m$_{v} \sim19\pm$0.5 mag. The approximate radio detection limit is indicated in Fig 4 where the flux density data have been plotted versus redshift. Note that this limit cannot be responsible for the upper envelope seen in Fig 1 since it only affects those sources that lie along the left-most edge of that plot.  In Fig 4, if the sources are good radio standard candles, then distance increases downward as indicated. Also, apparent optical brightness increases to lower redshifts, as shown by the apparent magnitudes included beside several of the sources. This is true regardless of whether one assumes the cosmological model (where redshift is interpreted as distance), or the local model (where most of the redshift is interpreted as intrinsic and related to the age of the object).

Also visible at the high redshift end of Fig 4 is an abrupt cut-off indicated by the dashed line. It is suggested here that this cut-off may be related to the optical plate cut-off mentioned above. In order to prove this it will be necessary a) to examine the magnitude distribution of the quasars and BL Lac objects in the \citet{sti94} sample to see if the distribution shows evidence of a cut-off at the early plate limit and b) to examine the optical magnitudes of the objects defining the cut-off in Fig 4. Fig 5 shows the distribution of magnitudes in the \citet{sti94} sample which contains 262 quasars and BL Lac objects. The histogram was obtained simply by counting the number of sources present in each half-magnitude-wide bin, and clearly shows that the sample of sources from which the present sample was selected has an abrupt magnitude cut-off at m$_{v} \sim19.5$ mag. The visual magnitudes of the 14 objects located along the dashed line in Fig 4, from \citet{kel98}, are listed in Table 1 where it is clear that they all have magnitudes that correspond closely to the early plate cut-off. These results strongly suggest that the cut-off in Fig 4 is indeed related to the plate cut-off (see also Section 4.5).
 
 The 14 sources that lie along the cut-off in Fig 4 are plotted as open circles in Fig 6 where the slope of the dashed line must now be explained. Recalling that if the sources are good standard candles the distance increases downwards, how can the more distant sources lying along the dashed line have the same apparent magnitude as the closer ones? Obviously this can only happen if the more distant ones are more luminous. In the DIR model sources become more luminous as they age and their intrinsic redshift \em decreases. \em The DIR model thus predicts that the dashed line in Figs 4 and 6 will have a \em positive \em slope as seen, if it is produced by the plate cut-off as suggested here.

In fact, it is possible to use the slope of the dashed line to determine how the quasar optical luminosity varies with intrinsic redshift (see below), at least for the intrinsic redshift range from z = $\sim0.8$ to $\sim4$. It is also worth noting that the upper envelope cut-off in Fig 1 is in no way related to the cut-off defined by the dashed line in Fig 4 because the sources that define these two cut-offs are completely different. Also, it is important to note again that this cut-off is defined solely by high-redshift quasars. If the sources are standard candles as suggested here, these high redshift sources are expected to be the most reliable distance indicators. It is also worth noting that, unlike in Fig 1, here the measurement uncertainties in redshift and flux density are much too small to affect the results.

Also apparent in Figs 4 and 6 is the fact that one source falls above the cut-off. This source (0153+744) has an apparent magnitude of 16$^{\rm m}$ and therefore would have been easily seen in early surveys. Although it has a measured redshift of z = 2.34, its intrinsic component must be much smaller for the source to be this luminous. Its measured redshift must then contain a much larger cosmological component than most other sources in this sample. The possibility that there might be other sources in this sample that fall into this same category cannot be ruled out.

\subsection{Variation of Optical Luminosity with Intrinsic Redshift}

Although born sub-luminous by several magnitudes, as a galaxy matures, and its intrinsic redshift decreases in the DIR model, its optical luminosity steadily increases as the surrounding host galaxy matures. If the central engine is a radio standard candle, and the 15 GHz radio flux density is then a measure of relative distance as claimed here, as the distance increases along the dashed line (downward) in Figs 4 and 6, the source luminosity must then increase by the same amount that the apparent brightness would normally be expected to decrease, if the optical apparent magnitude is to remain constant. Since, in the DIR model an increase in optical luminosity is accompanied by a decrease in intrinsic redshift, as noted above the slope of the cut-off in Fig 6 must be positive, exactly as found.

For an increase of a factor of ten in distance, the apparent magnitude m$_{v}$ must increase (source gets fainter) by 5 magnitudes. The luminosity must then $increase$ by 5 magnitudes if the apparent magnitude is to remain constant. From Fig 4, a factor of 10 increase in distance (a factor of 100 in flux density) is also associated with a factor of $\sim10$ decrease in the intrinsic redshift. Thus a decrease of a factor of $\sim10$ in intrinsic redshift implies an increase in luminosity of 5 magnitudes for the intrinsic redshift range between z = 1 and 3. Although no redshift corrections (ie K-corrections, etc.) have been applied to the data, this is not expected to affect the result because the range of redshifts covered is small (z = 1 to 2.5). Note that here z is assumed to be almost entirely intrinsic.

It is also of interest to compare this result to previous results obtained for the QSOs near NGC 1068. Four years ago a close examination of the 14 QSOs lying within 50\arcmin of the nearby Seyfert galaxy NGC 1068 found that 12 of these objects appeared to have been ejected from NGC 1068 in four similarly structured triplets, \citep{bel02a,bel02b,bel02c}. This picture allowed Doppler ejection components to be estimated, and since the cosmological redshift of NGC 1068 is known, for the first time accurate values for the intrinsic, or non-Doppler, components could be determined. When these results were combined with results obtained earlier by \citet{bur90}, this led to a well defined intrinsic redshift model that can now be tested. The relation defining the intrinsic redshifts in this model is given in \citet{bel03a} and \citet{bel04}. Fig 7 is a plot of data taken from Table 1 of \citet{bel02b} and shows that the optical luminosity was previously found to increase by 4 magnitudes when the intrinsic redshift decreased by a factor of 10, which is in reasonably good agreement with the value of 5 magnitudes found here. Again, no redshift corrections (k-corrections, etc.) have been applied, since the data cover a range of only two magnitudes. Although the results show good agreement, they are admittedly approximate and are for comparison purposes only.

Several other questions concerning the relations found using the data from \citet{kel04} are examined in Appendix A.

\begin{figure}
\hspace{-1.0cm}
\vspace{-2.0cm}
\epsscale{0.8}
\plotone{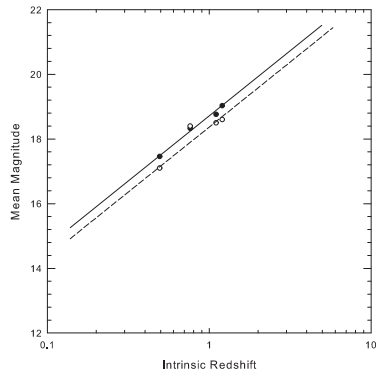}
\caption{{Optical magnitude change with intrinsic redshift obtained for sources near NGC 1068, taken from Table 1 of \citet{bel02b}. Filled circles are mean triplet values and open circles are mean pair values. Note that these are approximate values for comparison purposes only.  \label{fig7}}}
\end{figure}

\begin{figure}
\hspace{-1.0cm}
\vspace{-2.0cm}
\epsscale{1.0}
\plotone{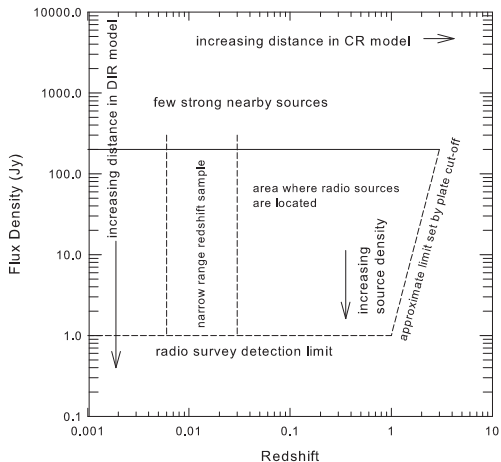}
\caption{{Flux density vs redshift plot showing area where radio sources are located. See Section 2 for an explanation of the slope of the upper redshift limit set by the plate limit. See text for an explanation of the two vertical dashed lines. \label{fig8}}}
\end{figure}

\subsection{On the source distribution in Fig 1}

If the blobs are ejected in completely random directions at similar velocities the distribution of sources in Fig 1 would be expected to be much more concentrated near the upper cut-off than is found. The dotted line in Fig 2 indicates the line above which 50 percent of the sources should fall if the distribution is random. Clearly this is not the case. However, this discrepancy is easily explained, at least in the DIR model, requiring only that the radiation from the central engine be directive, with the observed flux density increasing from edge-on to face-on as the region inside the hole in the torus becomes visible. This means that the flux density is only a measure of relative distance for a given inclination angle. However, this will not affect the slope of the upper envelope since these sources all have the same edge-on orientation. The sources in the bottom half of Fig 2 are nearer to face-on, thus if their measured flux density contains an extra component of radiation compared to the edge-on sources, the component attributable to their distance will be smaller than assumed here, and they will have moved into Fig 2 from the left by an amount equal to the additional contribution. In fact, the lower sources should be plotted much further to the left when their luminosity is normalized to the edge-on value and they cannot then be included in determining the source distribution with inclination. This is examined in more detail in Appendix B, and in \citet{bel06}, where it is demonstrated that when this effect is taken into account the source number distribution as a function of inclination angle closely resembles that expected for a random distribution of orientations. Since the distance component of the fluxes for the low-$\mu$ sources in Fig 2 will be lower, the distances for these sources in Fig 3 will be slightly greater. Although the flux density can vary with inclination angle, the total radiated output of these objects is still assumed to be a good radio standard candle and none of the conclusions reported here will be affected.

\section{Radio Characteristics of Source Samples in the CR and Local Models}

Here I look at two samples of extragalactic radio sources with jets, examining them in both the CR and DIR models.

\subsection{Radio Characteristics Expected in the CR and DIR Models}

Fig 8 shows, qualitatively, the area where most source flux densities are located when plotted versus redshift. In the CR model distance increases to the right. In the local model distance increases downward.

\begin{figure}
\hspace{-1.0cm}
\vspace{-1.7cm}
\epsscale{0.9}
\plotone{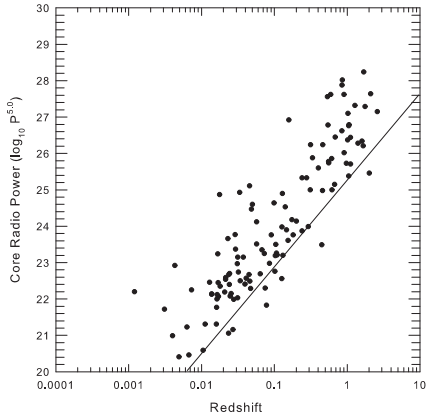}
\caption{{Radio power in the cores of 125 jetted sources assuming cosmological redshifts. Data are from \citet{bri84}. \label{fig9}}}
\end{figure}

\begin{figure}
\hspace{-1.0cm}
\vspace{-1.8cm}
\epsscale{1.0}
\plotone{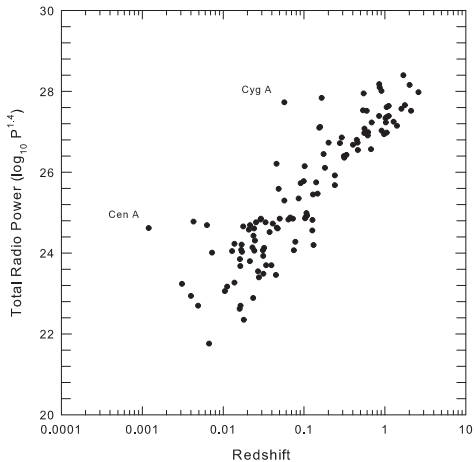}
\caption{{Total radio luminosity assumed to be mainly from radio lobes, assuming cosmological redshifts, plotted versus z. Data are from \citet{bri84}. \label{fig10}}}
\end{figure}

\begin{figure}
\hspace{-1.0cm}
\vspace{-1.8cm}
\epsscale{0.9}
\plotone{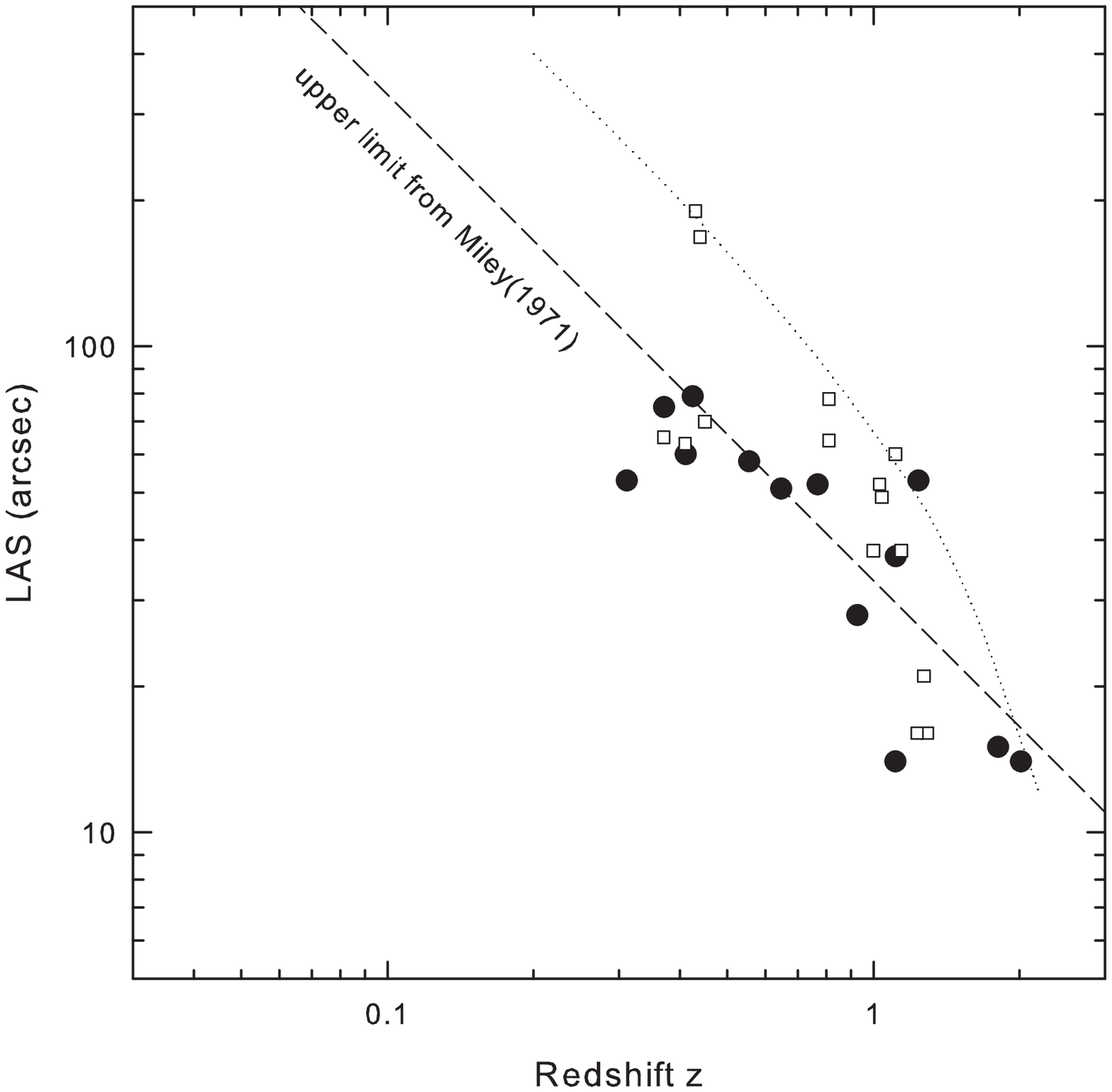}
\caption{{Largest angular separation of double radio lobes plotted versus z on logarithmic scales. (filled circle) from \citet{bri94}; (open squares) from \citet{goo03}; (dashed line)upper limit reported by \citet{mil71}; (dotted line) current upper envelope. \label{fig11}}}

\end{figure}

 
In the DIR model, although sources are born continuously throughout the age of the Universe, and objects with a complete range of intrinsic redshifts are expected to be present in all volumes of space \citep{bel04}, the active radio galaxies and quasars being examined here (that were detected in early radio surveys) are relatively nearby.  Their cosmological redshift components will then all be small compared to the intrinsic component. A slight decrease in the mean flux density at high intrinsic redshifts might be expected, but there can be no strong flux density fall-off with intrinsic redshift if the sources are standard candles. For a uniform space density, the sources with strong flux densities are expected to be few in number, simply because they are nearest, and the volume of space being sampled is small. This rapid decrease in the number of sources with decreasing space volume can be predicted to result in a relatively abrupt upper limit to the source distribution in Fig 8. Note also that in the DIR model, more distant radio sources will be found by improving the sensitivity of the radio finding survey.

\subsection{Radio Power in the Core}

In Fig 4 the total flux densities in the core and inner jets of 123 quasars, BL Lacs and radio galaxies measured at 15 GHz using the VLBA(on milli-arcsec scales) are plotted vs redshift. This is a raw data plot, where the data have been taken from \citet{kel04}, and it was shown in Section 2 above that these objects appear to be good radio standard candles.
There is little evidence that the average flux density at high redshifts is any smaller than at low redshift. In the CR model this requires that the source luminosity increase significantly with redshift, which in turn means that there is luminosity evolution. There must also be a wide range of luminosities at each redshift.

A second sample of radio sources with jets is also available in the literature. The 125 radio sources known in 1983 to have jets are listed in Table 1 of \citet{bri84}, where columns 3 and 4 list the core and total monochromatic powers respectively, after correcting for distance assuming the cosmological redshift model. The core power uses 5 GHz data and the total power uses 1.4 GHz data. In the latter, the total flux densities also include the contribution from the radio lobes. Only 10 of these jetted sources are common to the 123 sources plotted in Fig 4. If the sources are good standard candles, and the redshift is a measure of distance, plotting these radio luminosities versus redshift should result in a source distribution that is flat in the CR model. On the other hand, by converting the flux densities to radio luminosities by assuming that the intrinsic redshifts are related to distance when they are not (as is the case if the DIR model is correct), will produce a source distribution that has a steep positive slope.


In Fig 9 the radio power in the core of quasars and radio galaxies from \citet{bri84} is plotted versus the logarithm of the redshift. Note that this plot involves \em processed \em data, unlike the raw data plot in Fig 4. In the CR model the radio luminosity can be seen to increase steeply with increasing redshift, with an upper envelope slope close to that of the solid line. A similar result has been obtained by others \citep{hut88,nef89,nef90}, although the redshift range these investigators covered was much smaller, covering only the top half of the plot in Fig 9. In Fig 9 the solid line defines the approximate radio detection limit. In addition to the radio luminosity increase with redshift, several other conclusions can be drawn from this plot. Since the solid line in Fig 9 would have had a slope of zero before the conversion to luminosity was made, and the upper envelope of the source distribution is roughly parallel to this, it would also have had a slope near zero, and the source distribution would then have had the same appearance as in Fig 4. Furthermore, in Fig 9 there is no indication of where the active radio galaxies stop and the quasars begin. The plot is continuous, with no evidence of a discontinuity or a change in slope between quasars and active radio galaxies. In the CR model the radio luminosity of \em both \em the active radio galaxies and quasars must therefore increase steadily with increasing redshift \em with exactly the same slope. \em This argues strongly that if quasar redshifts are mainly intrinsic, \em so must be a large portion of the redshifts of most of the active galaxies in this plot. \em

Although these sources are defined only as galaxies and quasars, some of the objects with redshifts near 0.1 are likely to be BL Lac objects.  

In the DIR model, Fig 9 is exactly what would be predicted if the core flux densities are converted to luminosities assuming, incorrectly, that the redshifts are an indication of distance. In other words, the luminosity evolution that appears in the CR model is just that created by assuming that the intrinsic redshifts are cosmological.

\subsection{Radio Power in the Lobes}

The total radio luminosity (in the lobes, core, and jets) for the 125 sources in \citet{bri84}, is plotted versus the logarithm of the redshift in Fig 10. The mean total power is a factor of 40 stronger than that from the core alone. Since the jet component is assumed to be a relatively small fraction of this, I assume here that the total power is made up mainly of power from the lobes. However, since the core and total radio luminosities use data from 5 and 1.4 GHz respectively, some of this difference may be a spectral index effect. In Fig 10, as in Fig 9, the same large increase in radio luminosity with increasing redshift is seen in the CR model. The fact that the upper envelope is not as clearly defined in this plot may indicate that the radio lobes may be a poorer \em standard candle. \em However, the fact that the upper edge of both plots has a similar slope indicates that the luminosity increase with redshift in the core and lobes is quite similar. This suggests again that this is not a real increase in luminosity at all, but has instead been simply created by the assumption that the redshift is cosmological when it is not. Figs 9 and 10 resemble closely what would be expected if the local model is correct.


When the largest angular separation (LAS) between the two giant radio lobes of jetted sources is plotted vs redshift, the upper envelope of the source distribution is known to decrease as the redshift increases \citep{leg70,mil71}. In the CR model this decrease is mainly due to the increasing distance of the sources. However, even in those very early days, \citet{mil71} reported that the upper envelope appeared to fall off slightly more steeply than expected. Since the luminosity of both the core and lobes increases with redshift in the CR model, if anything, one might expect the linear separation to increase with increasing redshift. The density of the intergalactic medium may play a role, however \citep{mil71}. Fig 11 shows some more recent LAS values plotted vs redshift, using sources taken from \citet{bri94} and \citet{goo03}. Only sources that have clear evidence for a core and two lobes have been included to rule out doubles that might represent a core and one lobe. The dashed line represents the 1/z slope for Euclidean space taken from \citet{mil71}, and shows that the previous limit has been exceeded in many cases. There now appears to be even more evidence that the slope is steeper than expected, especially at high redshifts.

 In the DIR model the increase in LAS with \em decreasing \em intrinsic redshift is interpreted as a linear increase in the lobe separation as the source ages over a period of 10$^{7}$ to 10$^{8}$ yrs. In both models the LAS values that fall below the upper envelope are assumed to represent lobes that have been ejected closer to the line-of-sight. There is no problem explaining a steeper slope in the DIR model. Even a steepening at high redshifts, if real, could be explained by a less rapid initial decrease in intrinsic redshift.

\subsection{Intrinsic Redshifts in Radio Galaxies}

It has been argued \citep{bur04} that although there is significant evidence for intrinsic redshifts in quasars, the redshifts of low-redshift, $genuine$, radio galaxies are almost certainly purely cosmological. This question is important because, if it is true, the presence of apparent $superluminal$ motion in the jets of some of these low-redshift radio galaxies can be used to rule out the existence of all intrinsic redshifts. However, it has already been demonstrated that very small intrinsic redshift components are present in spiral galaxies \citep{tif96,tif97,bel02d,bel03a,bel03b,bel04a,rus05a,rus05b} and any additional links between the large intrinsic components in quasars and the very small ones seen in mature spiral galaxies would be significant information in the intrinsic redshift controversy. 

From Section 2 of this paper, a natural consequence of the sources being standard candles is that their radio distances (D$_{S}$) can be calculated. When these are known the intrinsic component can be obtained immediately, and it is quickly seen that some of these objects do contain a significant intrinsic component (see discussion on 3C120 below). Also, as discussed above, there is further confirmation of this in Fig 9, where the active radio galaxies and quasars are plotted using the same symbol it is clear that the upper envelope of the distribution is continuous over the entire redshift range for both the active radio galaxies and quasars.
As noted in the introduction, the intrinsic component in \em young, active \em radio galaxies is expected to be larger than that in \em mature \em radio galaxies, where a normal Hubble slope is seen (see also Section 4.5 below). The term \em genuine \em radio galaxy as discussed by \citet{bur04}, is thus assumed to apply to mature radio galaxies, and not the younger, active radio galaxies examined here. The easiest way to determine if a radio galaxy is young (still has a significant intrinsic component) is to determine its cosmological redshift component from its radio distance, and compare this to its measured redshift. An example of an active radio galaxy where superluminal motion has been claimed is 3C120, which has a redshift of z = 0.03 and $\beta_{app}$ = 4.6. If the distance to Virgo is 14.6 Mpc \citep{fre01}, the radio distance found here for 3C120 is z$_{c}$ = 0.0023 giving an intrinsic component of z$_{i}$ = 0.027 and $\beta_{app} \sim 0.35$.

\begin{figure}

\hspace{-1.0cm}
\vspace{-1.8cm}
\epsscale{1.0}
\plotone{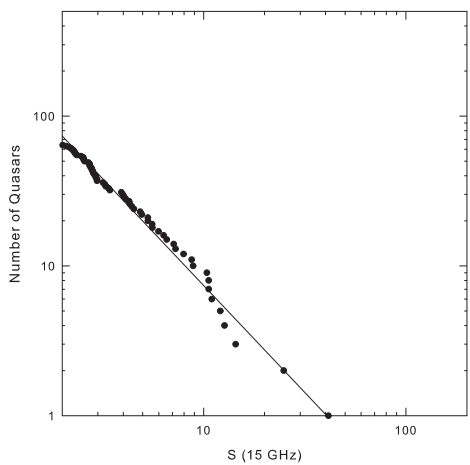}
\caption{{LogN-logS plot for sources with flux density S $>$ 2Jy, from \citet{kel04}. The slope of the line is -1.4. \label{fig12}}}

\end{figure}

\begin{figure}
\hspace{-1.0cm}
\vspace{-2.0cm}
\epsscale{1.0}
\plotone{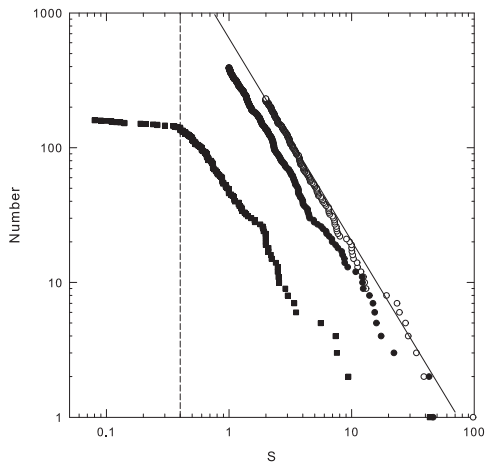}
\caption{{LogN-logS plot for sources from (square) \citet{wil78}, (open circle) \citet{wal85}, and (filled circle) \citet{sti94}. The solid line has a slope of -3/2. The dashed line indicates the approx. detection limit of the \citet{wil78} sample. \label{fig13}}}
\end{figure}

\section{Plots Involving Source Counts}

LogN-logS plots, that merely count the number of sources, N, to increasingly lower flux density levels, S, use raw data. No assumptions about the distance are involved.
If the sources are located in Euclidean space, as is indicated by Fig 3, and no evolution is involved (i.e. they are uniformly distributed standard candles) the logN-logS plot must then have a slope of -3/2. It is well known that for this slope to be obtained the objects do not have to be standard candles. Any luminosity distribution will produce the same slope so long as there is no evolution involved \citep{lon66}. However, the evidence presented in Section 2 indicates that in this case the objects \em are \em standard candles and are also located relatively nearby. Since no density evolution is likely within a 100 Mpc radius, a slope of -3/2 is predicted if the findings in Section 2 are correct. LogN-logS plots are then a test of the local model and this is examined below for four different source samples.

For standard candles, the relative distances, D$_{s}$, of the sources can also be obtained from the flux densities. For a uniform distribution in Euclidean space, logN-logD$_{s}$ plots must have a slope of +3.



\subsection{Data Samples for Source Count Plots}

The four quasar and active radio galaxy samples examined here are from \citet{wil78,wal85,sti94,kel04}. As discussed above, the \citet{kel04} sample contains 123, VLBA core flux densities measured at 15 GHz, and these values may represent the most accurate standard candles. However, it is assumed that the sources in the other samples will also be reasonably good standard radio candles, giving distances in the local model that are still much more accurate than the redshift distances. The \citet{wil78} sample contains flux densities for 160 quasars measured either at 178 MHz, 2700 MHz, or both. The \citet{wal85} all sky catalogue contains the 233 brightest radio sources at 2.7 GHz and is complete to 2 Jy over 9.81 sr of sky. The \citet{sti94} sample contains approximately 400 sources whose flux densities are measured at 5 GHz and exceed the detection limit of 1 Jy. Each sample has a clear radio detection limit visible, however, this is only a sharp limit in the \citet{sti94} and \citet{wal85} samples since the other two samples are less complete and, in some cases, involve flux densities obtained at a frequency other than that of the finding survey.

\subsection{LogN-logS plots}


Fig 12 is a logN-logS plot using the data in Fig 4. The solid line has a slope of -1.4, which is very close to that expected for a uniform source distribution if the sources are good standard candles in Euclidean space. Fig 13 is a similar plot of the data from the other three source samples. All plots again show good agreement with the slope of -3/2. It is also worth noting that the most complete samples \citep{wal85,sti94} give the best fits. This result allows us to conclude that the sources are uniformly distributed in the local model, as would be expected if the sources are nearby. In fact this result is quite remarkable, because it supports the local model without requiring any arbitrary assumptions.
This appears not to have been notice before, since previous investigations have only attempted to interpret the logN-logS slope in the CR model. In fact, finding it here is a direct result of the evidence found above that the sources are good standard candles. Although this slope of -3/2 can be explained in the CR model, it is unexpected because (a), there is luminosity evolution present, (b) there is a possibility that the space is not Euclidean, and (c), as will be demonstrated below, there is also evidence for strong density evolution in this model.

\subsection{LogN-logD$_{s}$ plots in the Local Model}

For standard candles, it is possible to use the flux densities to calculate the relative source distances, D$_{s}$, in the local model. When logN-logD$_{s}$ is plotted using these calculated distances, a slope of +3 is expected for a uniform distribution of radio standard candles in Euclidean space. This plot is shown in Fig 14 for the \citet{wil78} sample. The solid line has a slope of +3. Similar plots were obtained for the other source samples and, as for the logN-logS plots, the best fits are obtained for the most complete samples. In the local model this slope is exactly what is predicted and has resulted because the sources are uniformly distributed in this model. It is important to realize here that the uniform source distribution found \em comes from the raw data itself without requiring any additional assumptions. \em

\begin{figure}
\hspace{-1.0cm}
\vspace{-1.9cm}
\epsscale{1.0}
\plotone{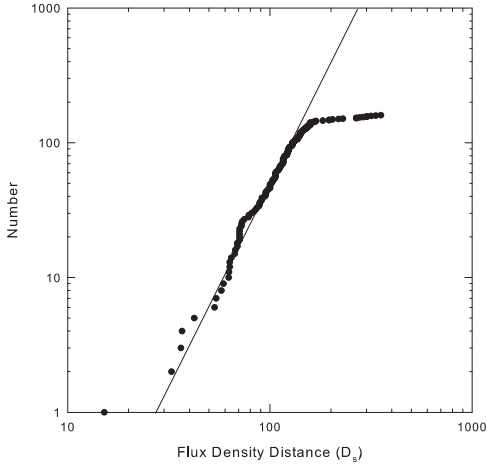}
\caption{{LogN-logD$_{s}$ plot for 160 sources from \citet{wil78}. D$_{s}$ is the relative distance calculated assuming that the sources are good radio standard candles. The line has the expected slope of +3. \label{fig14}}}
\end{figure}

\begin{figure}
\hspace{-1.0cm}
\vspace{-1.5cm}
\epsscale{1.0}
\plotone{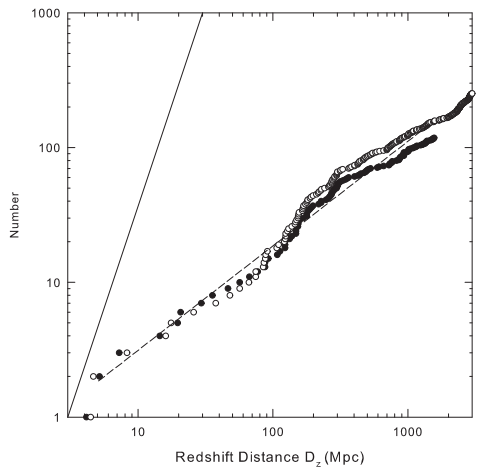}
\caption{{LogN-logD$_{z}$ plots for (filled circles) 231 sources from \citet{wal85}, and (open circles) 393 sources from \citet{sti94}. The dashed line has a slope of +0.77. \label{fig15}}}
\end{figure}

\subsection{LogN-logD$_{z}$ plots in the CR model}

In the cosmological redshift model it is possible to use the redshift to calculate an approximate redshift distance, D$_{z}$, and this has been calculated here from the relation D$_{z}$ (Mpc) =  (c/H$_{o}$)[(z+1)$^{2}$-1]/[(z+1)$^{2}$+1],

where c is the speed of light and H$_{o}$ is the Hubble constant. The value used for H$_{o}$ changes the distance obtained for all sources by the same factor and therefore cannot affect the slope. Fig 15 shows logN-logD$_{z}$ plots using redshift distances for the \citet{wal85} and \citet{sti94} samples. The dashed line gives the number vs redshift-distance slope, which is close to 0.8, and much less than the value of +3 that would be expected if the redshifts were a good measure of distance and the sources were uniformly distributed standard candles in Euclidean space. The radio luminosity was found to increase with redshift (Fig 9), and Fig 15 indicates that the source density must decrease significantly with increasing redshift. This is discussed in more detail in Sections 4.5 and 4.6 below where the reasons for this decrease are investigated. Similar results were obtained for the \citet{kel04} and \citet{wil78} samples.

\subsection{LogS vs Distance Plots in the CR model and the Generation of LogN-LogS plots}

In the CR model the source density was found above to decrease with increasing redshift faster than expected for a uniform distribution and the reason for this is now examined. Previously it has been assumed that this is due to the fact that more and more sources fall below the radio detection limit as the redshift increases (the Malmquist bias effect).

LogS-D$_{z}$ plots are shown in Figs 16, 17 18, and 19 for the four data samples. The radio detection limit is clearly visible in each figure. In the \citet{wil78} plot a few additional sources have been included by these authors, that have much lower flux densities. These sources are responsible for the change in slope in Figs 13 and 14 and do not indicate a real slope change. \citet{sti94} give a larger sample of flux densities at 5 GHz down to a limiting value of 1 Jy. The sample includes quasars, BL Lac objects and radio galaxies. It also includes many radio galaxies with redshifts above z = 0.3 that are not in the \citet{wal85} sample. The redshifts of these radio galaxies are expected to contain much larger cosmological components than the other radio galaxies in the samples being considered here, and these galaxies represent examples of the mature radio galaxies discussed above. It is clear from the logz-m$_{v}$ plot of these objects \citep[see Fig 7]{sti94} that the high redshift radio galaxies follow the expected Hubble slope, indicating that the largest component of their redshifts above z = 0.3 is cosmological. In the local model the source optical luminosity increases with age and therefore these older, more luminous, radio galaxies can be observed to much larger distances. They therefore cannot be considered to occupy the same volume of space as the sub-luminous (young) quasars, and hence cannot be part of the same sample of quasars, BL Lacs, and nearby active radio galaxies, which all have an intrinsic redshift component that is similar to, or larger than their cosmological component. Fig 19 includes all quasars and BL Lac objects with measured redshifts, but only those radio galaxies with redshifts below z = 0.3.

It is also of interest to note that the cut-off seen in the Kellermann sample at the high redshift end (Fig 16), and attributed to the presence of the early plate limit (see Section 2 above), is also visible in the source samples in Figs 17 and 18. It is less obvious in the Stickel sample (Fig 19) because an extra effort was made to obtain redshifts for sources whose apparent magnitudes were below (fainter than) the early plate cut-off. This is further proof that this cut-off is caused by the plate limit, since it disappears when the plate limit is removed. In the \citet{wil78} sample this high-redshift cutoff is visible down to a flux density level of S$_{2.7 GHz}$ = 0.1 Jy.

The plots in Figs 16, 17, 18, and 19 all resemble what is predicted in the local model (Fig 8) closely, with little evidence that the upper envelope falls off with increasing intrinsic redshift. Thus, in the CR model, the possibility that many more sources are lost at high redshift than at low redshift, because they have fallen below the radio detection limit (the Malmquist bias effect), can be ruled out. If sources are not lost because they fall below the detection limit, why is the slope of the logN-logD$_{z}$ plots in Fig 15 (0.8) so low? Also visible in the number vs distance plots in Figs 16, 17, 18 and 19, is the fact that none of them show the expected increase in the source number density with increasing redshift. This is then the main reason why the slope is 0.8 in Fig 15, instead of 3, and it means that a \em real \em decrease in spatial density with increasing redshift distance is required in the CR model. On the other hand there is an obvious increase in source density downwards in these figures in spite of the fact that the scale is logarithmic and expands downward. In order to be able to better compare the density change with distance in the two models, the logS scale has been converted to a linear distance scale in Fig 20. Here it is readily apparent by counting sources between dashed lines with equal separations, that the source density increases roughly as the cube when counting downwards, but remains relatively constant when counting to the right. This constant density with redshift (producing a slope near 1 in Fig 15) is exactly what is expected in the local model, where the redshifts are intrinsic, if the source density is uniformly distributed over the entire range of intrinsic redshifts.

\begin{figure}
\hspace{-1.0cm}
\vspace{-1.9cm}
\epsscale{1.0}
\plotone{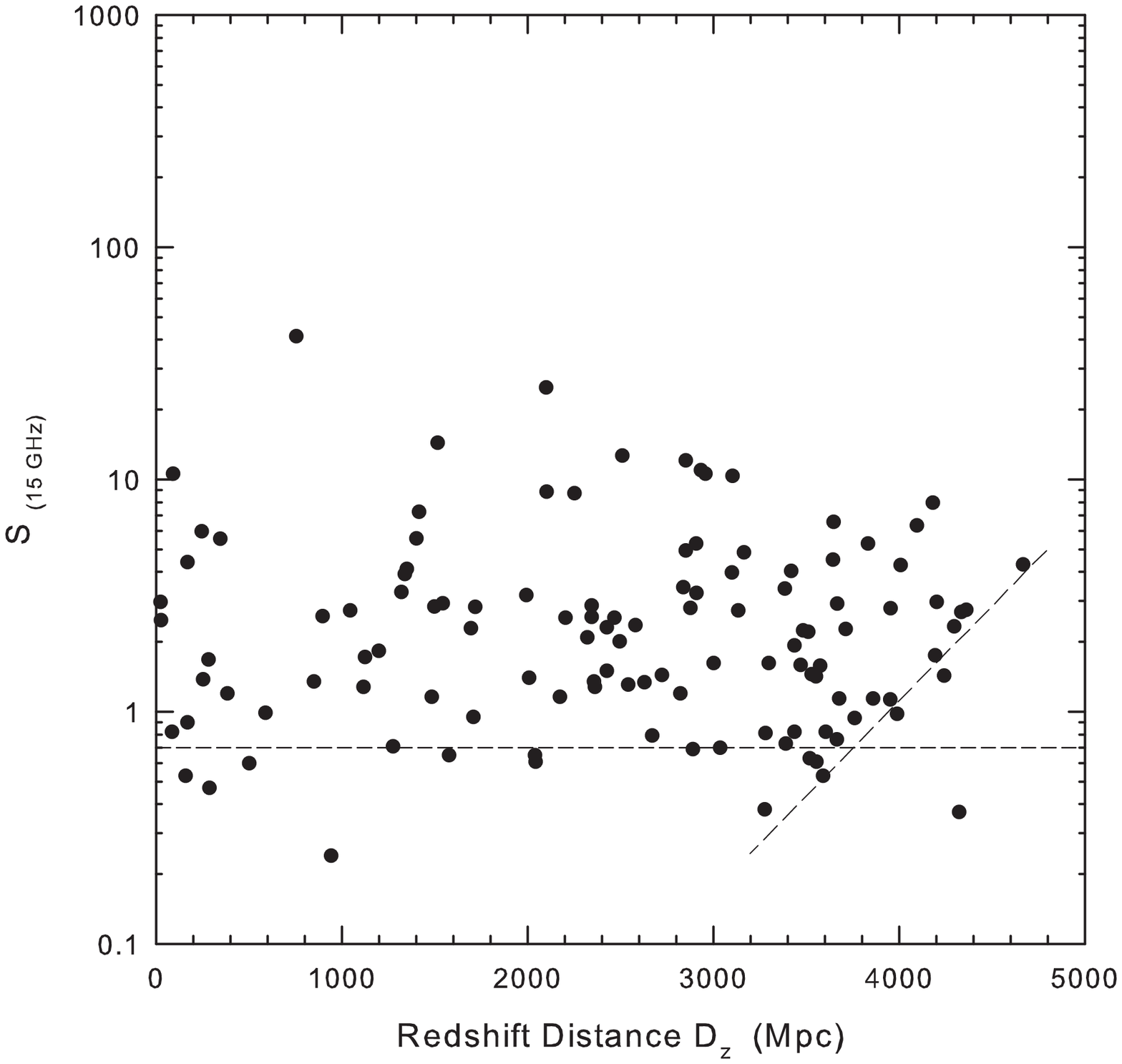}
\caption{{LogS plotted vs distance calculated from redshift from \citet{kel04}. \label{fig16}}}

\end{figure}

\begin{figure}
\hspace{-1.0cm}
\vspace{-1.9cm}
\epsscale{1.0}
\plotone{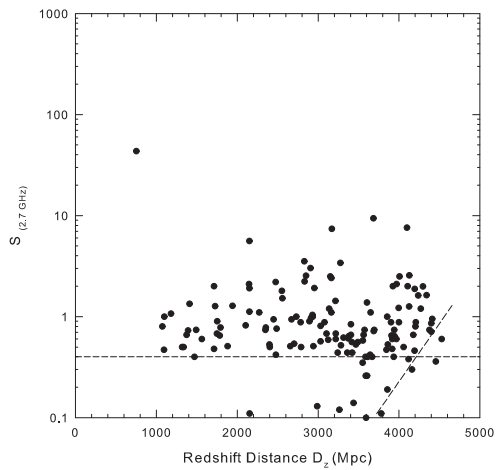}
\caption{{LogS plotted versus redshift distance for 142 sources with measured flux densities from \citet{wil78}. There is no evidence for an increase in source number density with increasing redshift. \label{fig17}}}

\end{figure}

\begin{figure}
\hspace{-1.0cm}
\vspace{-1.9cm}
\epsscale{1.0}
\plotone{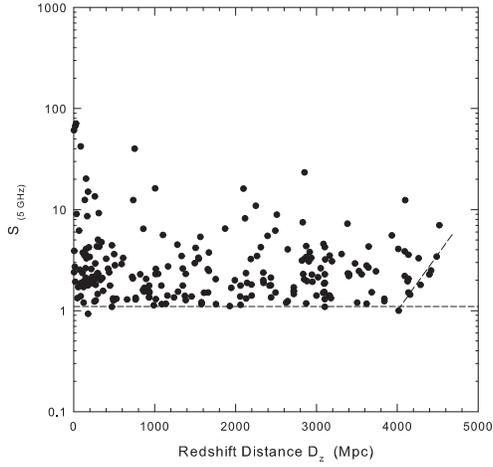}
\caption{{LogS$_{5 GHz}$ plotted versus redshift Distance D$_{z}$ for 231 sources from \citet{wal85}. \label{fig18}}}
\end{figure}

\begin{figure}
\hspace{-1.0cm}
\vspace{-2.0cm}
\epsscale{1.0}
\plotone{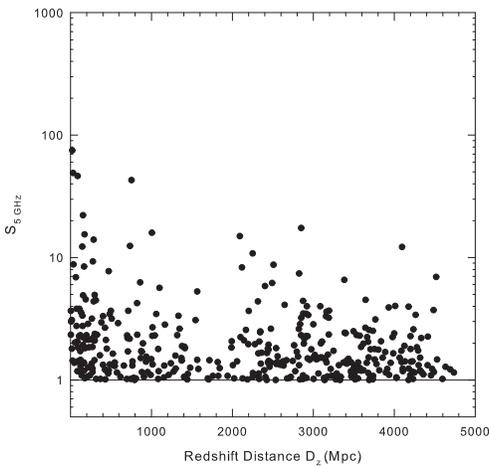}
\caption{{LogS$_{5 GHz}$ plotted versus redshift distance, D$_{z}$, for 393 sources from \citet{sti94}. \label{fig19}}}
\end{figure}

\begin{figure}
\hspace{-1.0cm}
\vspace{-1.5cm}
\epsscale{1.0}
\plotone{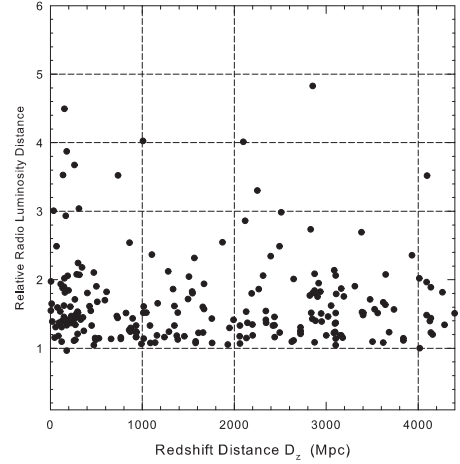}
\caption{{Relative Radio Distance plotted versus redshift Distance D$_{z}$ for 231 sources from \citet{wal85}. Distance increases downwards. See text for a description of the dashed lines.  \label{fig20}}}
\end{figure}

\begin{figure}
\hspace{-1.0cm}
\vspace{-1.9cm}
\epsscale{1.0}
\plotone{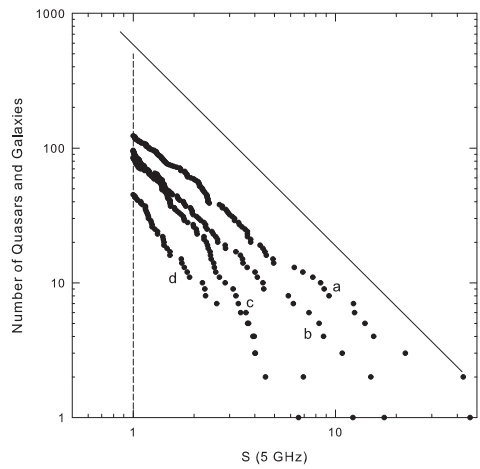}
\caption{{LogN-logS plots for sources in different distance ranges in Fig 19, where (a) 0-1000 Mpc, (b) 2000-3000 Mpc, (c) 3000-4000 Mpc, (d) 4000-5000 Mpc. Data from \citet{sti94}. \label{fig21}}}
\end{figure}

In the local model, if the redshift in the logS-D$_{z}$ plots (Figs 16 to 19) is divided into several narrow intrinsic redshift ranges (such as between the vertical dashed lines in Fig 8), the data in each is expected to produce a logN-logS slope close to -3/2.
In Fig 21, logN-logS plots are shown for sources that lie within different redshift-distance ranges for the \citet{sti94} sample. In Fig 21, (a) represents all sources with redshift distance between 0 and 1000 Mpc, (b) all sources between 2000 and 3000 Mpc, (c) between 3000 and 4000 Mpc and, (d) between 4000 and 5000 Mpc. These raw data plots show that there is good agreement with the expected logN-logS slope for uniformly distributed standard candles in each different redshift range. The variation in flux density in each narrow redshift range can be explained by a decrease in apparent radio brightness due to increasing source distance, and this change must be present if the local model is correct.
In the CR model this decrease in flux density is assumed to be due to a decrease in radio luminosity, which would then have to mimic the variation produced by the inverse square fall-off with distance seen in the local model. That this would happen by chance in the CR model seems rather unlikely.

\section{Interpretation of the Luminosity and Density Evolution Seen in the CR Model}

Although the luminosity and density evolutions required to fit the data in the CR model are arbitrary and without explanation in that model, the local DIR model predicts both exactly as found. In the local model the sources are uniformly distributed standard candles in Euclidean space and, because they are so close, their redshifts must be almost entirely intrinsic. If they are good standard candles, as found in Section 2, their luminosity \em cannot \em vary significantly with intrinsic redshift. This allows us to predict that the S vs redshift-distance distributions seen in Figs 16 to 19 will be flat, exactly as found. If it is assumed, \em incorrectly, \em that the intrinsic component is cosmological, the calculated radio luminosity will increase exactly as found in Fig 9. If sources are uniformly distributed with intrinsic redshift in the local model, the spatial density in the CR model will decrease as the cube of the distance, resulting in a slope close to 1 in the logN-logD$_{z}$ plot (Fig 15). That the slope found (0.8) is slightly different from the predicted slope of 1 is easily explained if either, or both of, the luminosity and spatial density are weak functions of intrinsic redshift.
The evolutions in luminosity and density in the CR model are thus predicted to be exactly as found, if the local model is correct.

\section{Discussion}

It has been demonstrated that the raw data fit the local DIR model perfectly, requiring neither luminosity nor spatial density evolution. When the data are fitted to the CR model, both luminosity and density evolution are required. These are difficult to interpret in the CR model because the luminosity increases with redshift while the density simultaneously decreases. However, as shown above, these evolutions in the CR model are predicted to be produced as found if intrinsic redshifts in the local model are assumed, incorrectly, to be cosmological. They are thus created solely by the processing carried out in an attempt to fit the data to the CR model. Although the data can be fitted to the CR model if enough assumptions are made, there is no real evidence from any of the raw data plots examined here that they support that model. However, when the logN-logS plots were first being examined 40 years ago, neither the DIR model, nor the fact that the sources could be radio standard candles, were known. This is no longer the case, however, and in Table 2 a comparison is made between the local and CR models using some of the evidence found here. The fact that the logN-logS plot is independent of distance is unfortunate. It is possible that astronomers may have been misled by these impressive plots, simply because they could still be explained in the CR model, even though the evidence presented here suggests that they are generated because the sources are uniformly distributed radio standard candles located in the very local universe.

As noted above, a few astronomers have continued to argue that high-redshift QSOs appear to have been ejected from nearby active galaxies, 
requiring that a large portion of their redshifts be unrelated to the Hubble expansion. And a significant amount of evidence has been reported claiming quasi-periodic quantization in the redshift distributions of quasars extending to normal galaxies. Until now, these two arguments have been the two main ones supporting intrinsic redshifts. When all the evidence presented here is considered together it strongly suggests that these objects are likely to be uniformly distributed standard radio candles in Euclidean space, and that they are much closer than their redshifts imply. These results then represent a third independent set of evidence supporting the claims that QSOs are born out of the nuclei of active galaxies, and that they initially have large non-Doppler redshift components that decrease with age.


\section{Conclusions}

Because new evidence of possible galaxy-quasar associations, and periodicities in redshift distributions, has continued to appear, the question of quasar distances has remained somewhat controversial for many years, and this question has been re-examined here using completely independent radio data. It was found, using angular motions in jetted sources and VLBA flux densities, that quasars and related active galaxies appear to be good radio standard candles. This is an important new result which leads to the conclusion that these objects are closer than 100 Mpc. This, in turn, allows several different predictions to be made, which can then be tested using radio data to see if this local model is internally consistent.
No density evolution is expected for sources so localized in space, and the redshift must be almost entirely intrinsic. For standard candles, no luminosity evolution is expected with intrinsic redshift, which predicts that flux density vs redshift plots will be flat. LogN-logD plots for complete samples of these objects must have a slope of 3, and logN-logS plots must have a slope of -3/2. All these predictions are shown to be filled, confirming that this local model is completely viable. If redshift is assumed to be a measure of distance, both the radio luminosity and space density of these objects are found to evolve with redshift. However, these evolutions turn out to be just those predicted in the local model when intrinsic redshifts are assumed, incorrectly, to be a measure of distance. Because the local model provides an excellent fit in all instances to the simplest model, it is concluded that flux density, and not redshift, is likely to be the best indication of the distance to these objects. If so they must be a lot closer than their redshifts imply.

\section{Acknowledgements}

I thank Dr. D. McDiarmid for many helpful comments, and Simon Comeau for assistance with the data preparation.

\section{Appendix A}

\subsection{Can the upper envelope in Fig 1 have occurred by chance?}

To investigate the possibility that the upper envelope in Fig 1 might have occurred by chance several test datasets were generated. To keep the test datasets as similar to the real one as possible they were generated using the following procedure. The two columns containing the source angular motions and 15 GHz flux densities were first placed side by side. A new dataset was generated by shifting one of the columns relative to the other by one channel. Each additional shift of one channel produced a new dataset. This was first repeated for six one-channel shifts and finally, for a shift involving half of the data. None of the generated datasets produced plots with a clear upper envelope, let alone one with the correct slope. It was concluded that it is extremely unlikely that the upper envelope seen in Fig 1 was produced by chance.

\subsection{Are the VLBA flux densities good standard candles?}

Because of its importance to the results obtained here this question is now considered in more detail. The 15 GHz flux densities used in this analysis are described by \cite{kel04} as the $largest$ total flux density seen on any of the VLBA images at 15 GHz. The source sample includes all known sources in the \citet{sti94} catalogue that have a flat spectral index anywhere above 500 MHz. This source sample is a flux density limited sub-sample of the full 15 GHz VLBA survey, and was obtained by using the measured VLBA flux densities as the main selection criterion. All flat spectrum sources that had a total CLEAN VLBA flux density exceeding 1.5 Jy (2 Jy for southern sources) at any epoch since 1994 are included in this sub-sample. The flux densities came from measurements made over extended periods up to 7 years in duration and many of the sources will have been variable over this period, with outbursts typically lasting the order of one year. Fractional flux density changes in variable sources range from a few percent to about 100 percent \citep[page 581]{ver88}. Because the slope is $\onehalf$ in Fig 1, the flux density variations have to be twice as large to produce the same effect as the errors in angular motion. Clearly, even a few sources with flux densities in error by a factor of two will not affect the slope in Fig 1 where the total range in flux density covers more than a factor of 50. It is therefore concluded that although some of these sources are variables they can still be considered good standard candles for the purposes of this investigation. There may also be some evidence that the largest variations occur in radio galaxies which have not been included in Fig 1, and are not associated with the cut-off in Fig 4. Is there some reason to suspect that the VLBA total flux density might be a good distance indicator? On the VLBA images (fig 2 of Kellermann et al 1998) the core component, where the outbursts occur, dominates in almost all cases. Since the flux density used here is the $largest$ total flux seen on any of the images during the monitoring period, one might also speculate that the outburst itself is a standard candle, resembling the standard candle characteristics of SnIa peak luminosities. If so, the flux densities obtained in this investigation would represent a unique set of data. Perhaps the best evidence that the VLBA flux density is a good distance indicator is the sharpness of the upper envelope in Fig 1 and of the cut-off in Figs 4 and 6. The 25 sources best known as variables from early work \citep[Table 1]{kel68} are part of the present sample. However, none of these is amongst the 14 high-redshift quasars defining the cut-off in Figs 4 and 6.

Why has it not been noticed previously that these sources are good radio standard candles? Even using these VLBA flux densities, had they been previously available, it would not have been noticed that they were a good measure of relative distance because any attempt to detect this would have been carried out using the cosmological model, with the measured redshift as the distance parameter. As can be seen from Fig 4, if redshifts are assumed to be cosmological the flux densities do not show the expected decrease with increasing redshift that would be expected if the sources were standard candles. In fact, this is a good example of the danger involved in not giving fair consideration to all available evidence on both sides of contentious issues before conclusions are reached. This is especially true here, where the nature of the redshifts is a question that is as fundamental to astronomy as the Hubble constant itself.


\subsection{Can the cut-off in Fig 4 be explained in the cosmological redshift model?} 

The high-z cut-off in Fig 4 is clearly visible and, although it can easily be explained in the local model, the possibility that it might also be explained in the cosmological model needs to be examined. In a radio finding survey with a fixed detection limit, the limit cannot depart from this value as the redshift increases. However, since the radio limit of 1 Jy was set in the 5 GHz survey, a departure could occur when fluxes are measured for these same sources at a different frequency. However, this is unlikely to be the explanation for the cut-off in Fig 4 since it would require an abrupt and \em systematic \em change in spectral index between 5 and 15 GHz, starting at z $\sim1$. It would also require that the sources falling along the cut-off in Fig 4 at 15 GHz be the same sources that fall along the 1 Jy cut-off at 5 GHz. This is easily checked and it is found that the sources that lie along the cut-off in Fig 4 do not have flux densities of 1 Jy at 5 GHz. This indicates that some effect other than a systematic change in spectral index must have produced the cut-off in Fig 4.

Another possibility that needs to be considered is whether or not the early plate cut-off, which is clearly present in the 5 GHz data in Fig 5, could also explain the cut-off in Fig 4 in the CR model. Although the plate cut-off cannot affect the original radio limit, it presumably can affect the final source list, since the redshifts of these sources could not be measured until the sources were identified on photographic plates. Furthermore the few sources that fall above the plate cut-off in Fig 5 would have had to have been observed to higher optical sensitivity and perhaps should not even be considered with the main sample that may otherwise be complete to the early plate limit. However, in the DIR model the visibility of the plate cut-off can be explained because a) the radio flux density is a measure of distance and b) the source luminosity decreases with increasing intrinsic redshift. In the cosmological model the apparent magnitudes also get fainter with increasing redshift, but there is no evidence in this model that the radio flux density is a standard candle. Without this, when plotted as a function of the flux density there appears to be no way that the plate cut-off could be aligned along the cut-off seen in Figs 4 and 6.

There thus appears to be no simple explanation for the cut-off in Fig 4 in the CR model. It is conceivable that complex flux related arguments might be offered to explain this relation since it is a function of the flux density, however, since the total flux is measured, and the un-beamed core component appears to be the dominant component on the VLBA maps at all redshifts, arguments that involve selective Doppler boosting in the jet at high redshifts are also unlikely to be able to explain this result.


 It is also worth noting that the cut-off in Fig 4 has the same slope regardless of whether one uses the total 15 GHz flux from \citet{kel04}, the total 15 GHz flux from \citet{kel98}, or the Jy/beam 15 GHz flux value from \citet{kel98}.
This cut-off is also clearly visible in other source samples (see Figs 17 and 18). It is not visible in the Stickel sample because a special effort was made to obtain redshifts of sources in that sample that were beyond (fainter than) the old plate limit.

\subsection{Can the upper envelope cut-off in Fig 1 be explained in the cosmological model?}

In the relativistic beaming model the sources with the fastest apparent motion will be those in which the jet is aligned most closely with the line-of-sight, with $\mu(1+z)$ proportional to the \em blob \em Lorentz factor $\gamma$. Although the flux density from a \em blob \em may be related to its $\gamma$, this flux is not related closely to the total flux plotted in Fig 1. In Fig 2 of Kellermann et al. (1998) where the maps of the individual sources are presented, the core flux density dominates that from an individual \em blob \em by an enormous amount in all cases. Furthermore, the flux density from the core varies significantly with no correlated variation seen in the jet, indicating that the flux density of a particular \em blob \em, or even the entire jet, is unlikely to be significantly correlated with the total (core + inner jet) flux density plotted in Fig 1. There appears to be no reason then that there should be any relation between the total flux and the angular motion, let alone one with a slope of one-half. Although there are many other corrections and assumptions that need to be taken into account in the cosmological model it is hard to believe that after making these, the tight angular motion vs flux density relation seen along the upper envelope in Fig 1 would be any more explainable. Furthermore, the results obtained after making many corrections and assumptions cannot approach the significance of results obtained in raw data plots where none of these corrections are required. One has only to look at a small fraction of the papers published in this field to see that even with the enormous efforts that have been made to date, there are still many unanswered questions about superluminal sources in the cosmological redshift model. Furthermore, even obtaining limited agreement with the data in that model has only come at the price of making the interpretation much more complex.

\section{Appendix B}

On the Source Distribution in Figs 1 and 2.

If it is assumed that the blobs are all ejected along the rotation axis with reasonably similar velocities, the projected velocity v$_{p}$, which is proportional to the angle between the ejection direction and the l-o-s, is also, therefore, proportional to the inclination angle $i$ of the central torus. For similar ejection speeds the projected ejection velocity v$_{p}$, which is proportional to the angular motion $\mu$ (mas yr$^{-1}$) of the blobs in the inner jets, will be dependent on the ejection angle $i$. The observed projected ejection velocity can then be converted to the inclination angle of the central torus using the relation v$_{p}$ = v$_{e}$sin$i$, where v$_{e}$ is the ejection speed.

By normalizing the $\mu$ of all sources to the same distance using the flux density, the distribution of sources versus inclination angle can be obtained. This has been plotted in Fig 22 after binning the sources into $3\arcdeg$-wide bins. It can be seen that the observed distribution falls off almost exponentially with inclination angle. This is contrary to the shape of the distribution expected for random ejections when no selection effects are present, which is represented by the dashed line and reveals that for a random distribution of orientations, $50\%$ of the sources are expected to be located above $60\arcdeg$. As noted above, the dotted line in Fig 2 also represents the $50\%$ line. This discrepancy needs to be explained, regardless of which redshift model (cosmological or intrinsic) is assumed. In the cosmological redshift model Doppler boosting is a possible explanation, although it is then difficult to explain the upper envelope and slope of 0.5 in Fig 2.



\begin{figure}
\hspace{-1.0cm}
\vspace{-1.0cm}
\epsscale{0.9}
\plotone{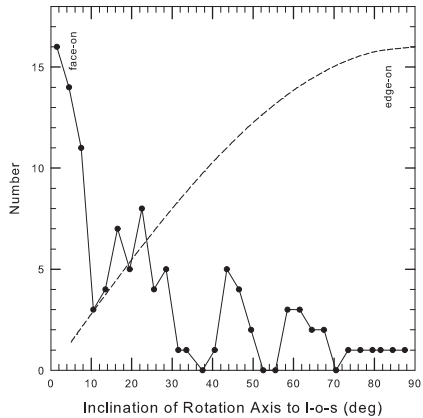}
\caption{{(solid line)Distribution of sources as a function of the torus inclination angle after binning into $3\arcdeg$-wide bins. (dashed line) distribution expected for a random orientations with no selection effects. \label{fig22}}}
\end{figure}

\subsection{Source Detection as a Function of Radio Flux Density}

In the DIR model I claim here that the sources are radio standard candles for a given inclination angle and the radio flux density is then also a measure of source relative distance for sources at that inclination angle. For a flux limited sample an increase in radio luminosity, or a lowering of the detection limit, will translate into a large increase in the number of detected sources because the number of sources goes as the cube of the distance. The increase in source density in Fig 22 as the inclination angle, $i$, decreases towards face-on, which is counter to that expected if the sources are standard candles, can be explained by an increase in the radiated flux as the inclination angle decreases. In other words, the luminosity of all sources would be the same for any specified inclination angle but it would vary with that angle, becoming strongest when the object is viewed directly into the central hole in the torus. In Fig 2, this means that the source luminosity increases as one moves downward from the solid line because the inclination angle is simultaneously decreasing.  If the luminosity of each source was normalized to that for $i$ = $90\arcdeg$, the location of any given source below the line in Fig 2 would move to the left horizontally, by an amount that depends on its present distance below the upper envelope.

What could be the origin of such a luminosity dependence? One possibility is that the radiation comes from an optically thick torus and there is a similar decrease in area of the torus from face-on to edge-on. In the DIR model a factor of 4 decrease in flux density is equivalent to a factor 2 increase in distance. For a flux limited sample, and since the number of sources increases as the cube of the distance, a factor of 4 decrease in flux density (visible area of the torus), between face-on and edge-on, will translate into a factor of 8 decrease in number for edge-on sources detected. Fig 23 shows an example of an edge-on and a face-on torus. The relative decrease in area from face-on to edge on will increase as the torus flattening increases and the dashed lines in Fig 23 move closer together. This then is a selection effect that can explain at least some of the decrease seen in the number of sources near the upper envelope in Fig 1. However, large area changes will require significant flattening of the torus, presumably due to rotation, and this method alone may not be able to explain the large decrease seen with increasing inclinaton angle in Fig 22.

\begin{figure}
\hspace{-1.0cm}
\vspace{-1.0cm}
\epsscale{1.0}
\plotone{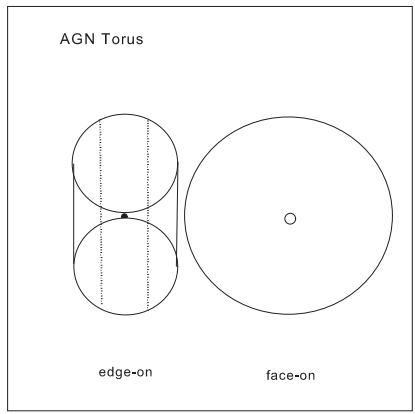}
\caption{{Change in area between edge-on and face-on optically thick torus. \label{fig23}}}
\end{figure}

\begin{figure}
\hspace{-1.0cm}
\vspace{-1.6cm}
\epsscale{0.9}
\plotone{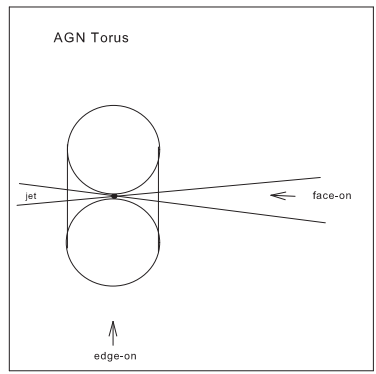}
\caption{{Torus with two-sided jet viewed edge-on. \label{fig24}}}
\end{figure}

The overall shape of the number vs inclination angle plot in Fig 22 suggests an additional explanation for the decrease in flux density between face-on and edge-on sources. Since the fall-off in source number begins within a few degrees of the face-on position, there may be a component of the observed flux density that is unrelated to radiation from the torus. It may be coming from inside the hole in the torus where the ejected blobs originate. In this case its strength would be entirely related to the visibility of this inner region which, in turn, would be controlled by the inclination angle of the torus, as demonstrated in Fig 24. In fact, it may be possible to obtain an estimate of the opening angle of this radiation emerging from the hole in the torus using the number distribution vs inclination angle. In this scenario there is no problem producing a large change in observed flux between face-on and edge-on inclinations.

\begin{figure}
\hspace{-1.0cm}
\vspace{-1.0cm}
\epsscale{1.0}
\plotone{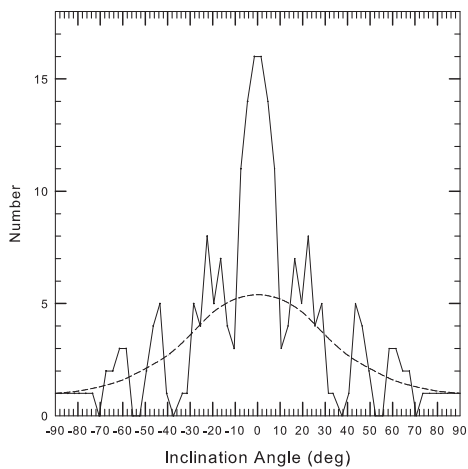}
\caption{{Change in observed source density with torus rotation. \label{fig25}}}
\end{figure}

\begin{figure}
\hspace{-1.0cm}
\vspace{-1.6cm}
\epsscale{0.9}
\plotone{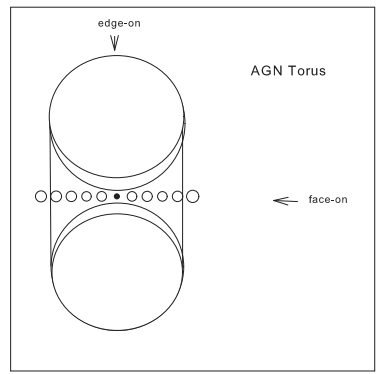}
\caption{{Change in radio flux density between edge-on and face-on orientations of optically thick torus. \label{fig26}}}
\end{figure}

In Fig 25 the observed distribution in Fig 22 has been re-plotted assuming that negative rotation angles would result in a mirror image of what is seen for positive rotations. This resulting distribution is remarkably similar to that of an antenna radiation pattern. However, it is unlikely that any pattern (including sidelobes) could be produced in the conventional manner. Emission may come from a combination of sources, including material near the central black hole, material throughout the hole in the torus, and material along the inner crescent on the inside edge of the torus as shown in Fig 26. However, after further examination of Fig 25 it is concluded, based on the shape of the broad curve designated by a dashed line, that this component may be produced by a fat, optically thick, torus as its orientation changes. This number density variation near a factor of 5 can easily be explained by the area change. It is further concluded that the strong narrow feature within $\sim10\arcdeg$ of face-on is produced by much stronger radiation emerging from the hole in the torus. This conclusion appears necessary since it appears to be too strong to be explained by a change in area, and because a change in area could not explain the sudden fall-off within a few degrees of face-on, although this fall-off is unlikely to be as steep as indicated in Fig 25 (see \citet{bel06} for a more complete discussion). The decrease in source density seen in Fig 25 can thus be explained by a simultaneous decrease in flux density as the orientation changes from face-on ($i = 0\arcdeg$) to edge-on ($i = 90\arcdeg$). 

As noted above, source flux at 90$\arcdeg$ is a measure of distance in this model. Thus the number of sources will vary as S(90$\arcdeg)^{2/3}$. If S is not dependent on $i$, the source number density as a function of $i$ will vary as sin$i$. However, if S is $i$-dependent, we obtain another distribution, N($i$), owing to the fact that the distance corresponding to S(90$\arcdeg$) is different from that for S($i$). The difference is proportional to (S($i$)/S(90$\arcdeg))^{3/2}$.  This means that

S($i$)/S(90$\arcdeg$)  =  [N($i$)/sin$i]^{2/3}$  --- (1)

The conversion of the solid line in Fig 25 to S($i$)/S(90$\arcdeg$) is shown by the solid line in Fig 27.  This, however, is an approximation because the determination of the inclination angles significantly below 90$\arcdeg$ is an approximation as is now shown.

\subsection{Correction of the Observed Flux Density and Distance}

Fig 27 shows that S($i$)/S(90$\arcdeg$) peaks near $i$ = 0$\arcdeg$, which means that if the S values for all of the sources are normalized to S(90$\arcdeg$) to give us a uniform measure of distance, the points below the line in Fig 2 will move to the left, especially those well below the line whose inclination angle is small.  A new determination of the inclination angles based on these new source positions will yield more accurate values. Because the movement to the left results in each point becoming closer to the line above, the inclination angles are increased.  The increase is very small for sources whose $i$ is near $90\arcdeg$ and zero for those with $i$ = 90$\arcdeg$ (which means that the slope of the upper envelope will not be affected).  On the other hand, the increase is large for those sources whose $i \ll 90\arcdeg$ because they must be moved significantly to the left as indicated by Fig 27.  Consequently the width of the peak in Fig 27 is actually larger than shown, as we will see below.


\begin{figure}
\hspace{-1.0cm}
\vspace{-1.6cm}
\epsscale{0.9}
\plotone{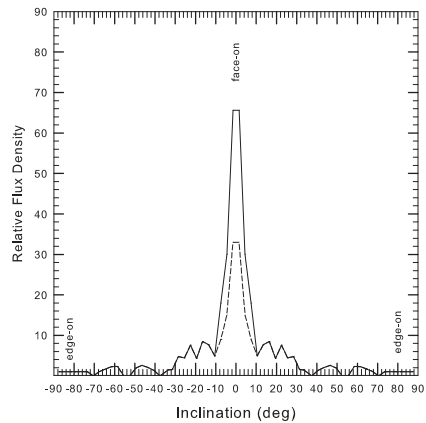}
\caption{{Change in flux density required with torus rotation to produce density distribution in Fig. 25. See text for a description of the dashed curve. \label{fig27}}}
\end{figure}

\begin{figure}
\hspace{-1.0cm}
\vspace{-1.0cm}
\epsscale{1.0}
\plotone{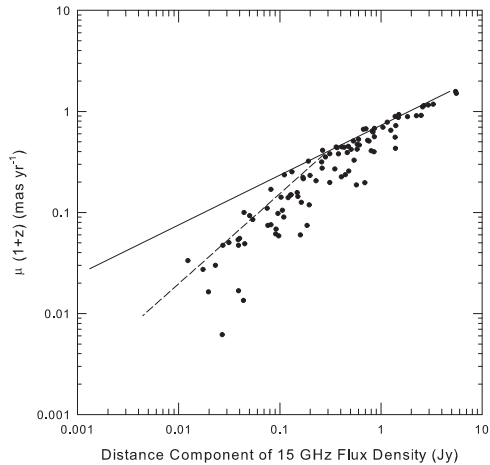}
\caption{{Fig 1 after the flux densities were corrected using the solid curve in Fig 27. Dashed line represents the approximate detection limit. \label{fig28}}}
\end{figure}

For clarity, I describe this analytical process again. If the true flux density change with inclination angle is known, it can be used to normalize all observed flux density values to one inclination angle. Correcting the fluxes to edge-on by removing the intrinsic flux density component will move sources to the left in Fig 2, with the amount of movement proportional to their distance below the upper envelope. Put another way, many of the sources that lie well below the upper envelope in Fig 2 will have been plotted in their present location simply because they contain a large intrinsic flux component. Not only will they have originated in a distant region where there are many more sources to choose from because of the increased volume of space sampled, they are clearly not from the same space in which those near the upper envelope in Fig 1 are located, since these cannot have moved significantly. They should then not be included in the same number distribution count.

 In Fig 28 the solid curve in Fig 27 has been used to normalize the intrinsic component of the flux at all inclinations to the edge-on value, so that relative distances can be compared. For distances greater than that corresponding to S $\sim 0.2$ Jy the source number density near edge-on will have been reduced because of the detection limit, which is shown approximately by the dashed line. Because the inclination angle estimated for each source in Fig 2 will be too small, the width of the peak near face-on in Fig 27 will be too narrow. These under-estimated inclination angles will also have led to an over-correction in Fig 28. Thus Fig 28 is assumed to represent an upper limit to the adjustment required to move the sources to their original distances.

 \begin{figure}
\hspace{-1.0cm}
\vspace{-1.6cm}
\epsscale{1.0}
\plotone{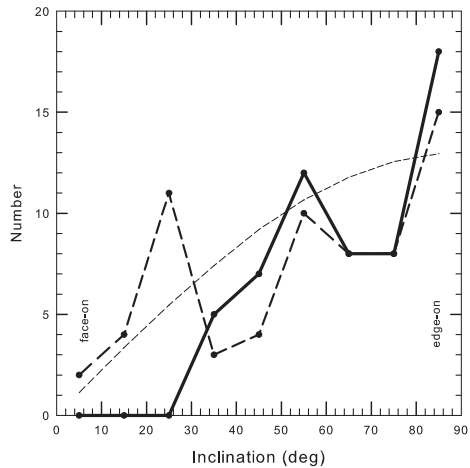}
\caption{{Number distribution for sources with corrected flux densities greater than S = 0.2 Jy as a function of inclination angle after binning into $10\arcdeg$ bins as described in the text. \label{fig29}}}
\end{figure}
  


In Fig 29 the number density distribution has been recalculated (solid line) using only those sources above S = 0.2 Jy in Fig 28 to avoid the area where the number has been affected by the detection limit. This curve is now a good fit to the expected curve shown by the dashed line. When the dashed curve in Fig 27 was used to correct the flux density, the number density distribution shown by the bold dashed line in Fig 29 was obtained. The dashed curve in Fig 27 was obtained simply by scaling down the values near the central peak by a factor of 2 in an attempt to correct for the underestimated inclination angles discussed above.

It is concluded from this analysis that the source distribution in Fig 2 can be explained if there is a reasonable increase in radio luminosity when the inclination angle of the central torus approaches face-on. A more detailed discussion of this is presented in \citet{bel06} where an estimate of the opening angle of the hole in the torus is also made.

\clearpage








\begin{deluxetable}{ccccc}
\tabletypesize{\scriptsize}

\tablecaption{Sources at the Plate Cut-off. \label{tbl-1}}
\tablewidth{0pt}
\tablehead{
\colhead{Source} & \colhead{S$_{VLBI}$} & \colhead{z} & \colhead{$\mu$(mas yr$^{-1}$)} & \colhead{m$_{\rm v}$}

}

\startdata

0016+731 & 0.98 & 1.781 & 0.074  & 18.0 \\

0035+413 & 0.53 & 1.353 & 0.1 &  19.1 \\
0212+735 & 2.69 & 2.367 & 0.08 &  19.0  \\
0458-020 & 2.33 & 2.286 & 0.1 &  18.4 \\

0642+449 & 4.31 & 3.408 & 0.01 &  18.5 \\
0707+476 & 0.63 & 1.292 & 0.11 &  18.2 \\
0850+581 & 0.61 & 1.322 & 0.2 &  18.0 \\

0917+449 & 1.43 & 2.18 & 0.07 &  19.0  \\
1128+385 & 1.13 & 1.733 & 0.01 &  18.0  \\

1532+016 & 0.76 & 1.42 & 0.21 &  18.7 \\

1656+477 & 1.14 & 1.622 & 0.06 &  18.0  \\ 
1758+388 & 1.75 & 2.092 & 0.002 &  18.0  \\

2113+293 & 0.94 & 1.514 & 0.02 &  19.5 \\
2136+141 & 2.75 & 2.472 & 0.02 &  18.5 \\

\enddata 

\end{deluxetable}

\begin{deluxetable}{lll}
\tabletypesize{\scriptsize}
\tablecaption{Comparison Between the Local and CR Models. \label{tbl-2}}
\tablewidth{0pt}
\tablehead{

\colhead{ } & \colhead{Local Model} & \colhead{CR Model}
}

\startdata

1) Can the observations be explained without relativistic velocities? & Yes & No \\
2) Does the logN-logD plot have the correct slope of three? & Yes & No \\
3) Can the observations be explained without having to assume an arbitrary luminosity evolution?  & Yes  & No \\
4) Can the observations be fitted without assuming an arbitrary density evolution?  & Yes  & No  \\
5) Is there corroborating evidence that the sources are standard candles? &  Yes  &  No  \\
6) Is the logN-logS plot generated in the simplest manner predicted for the model?  &  Yes  & No  \\

\enddata 

\end{deluxetable}

\end{document}